\documentclass[prl,twocolumn,amsmath,amssymb]{revtex4}

\usepackage{graphicx}
\usepackage{bm}
\usepackage[usenames,dvipsnames]{color}
\definecolor{altncolor}{rgb}{0,0,0.8}
\usepackage[colorlinks=true, linkcolor=green, anchorcolor=altncolor,
citecolor=altncolor, filecolor=altncolor, menucolor=altncolor,
urlcolor=altncolor]{hyperref}

\begin{document}

\title{Entangled photon detection and ephemeral space-like Schr{\"o}dinger cat states}

\author{Peter B. Weichman}

\affiliation{BAE Systems, FAST Labs, 600 District Avenue, Burlington, MA 01803}




\begin{abstract}

A model of single photon detection, illustrated by a photon-absorbing superfluid or superconducting microvolume, is formulated as a cascading pair of quantum phase transitions. In the first, the microvolume transitions to the normal state upon photon absorption, resulting in a superposition of macrostates depending on whether the photon is absorbed or not. The second enables subsequent ``wavefunction collapse,'' producing a density matrix implementing the Born probability rule. Next, EPR-type measurements on space-like separated entangled photon pairs are considered. It is argued that macro-entangled superposition indeed survives until such time as the component states come into causal contact, following which the state rapidly collapses to one or the other expected outcome. Apparent superluminal communication effects are entirely avoided.

\end{abstract}

\maketitle

We construct and analyze here an idealized but quantitative model of single photon detection, with the aim of deriving predictions for measurement outcomes for single and multiple entangled photon detection entirely within the context of the many body Schr\"odinger equation. We provide here a high level overview, though the model should be tractable enough to support future more detailed analytic and numerical support for the conclusions.

Individual detectors consist of Bose superfluid microsamples (which could also be interpreted as a system of superconducting Cooper pairs), beginning at some ordered phase temperature $T < T_c$, coupled to the electromagnetic (EM) background field. The samples are macroscopically small, but with sufficiently large particle number $N_B$ that their equilibria may still be treated in the thermodynamic limit on measurement time scales. We consider first the case in which a single photon pulse propagates through the detector volume. The photon may be absorbed, partially absorbed, or scattered, each with some amplitude, or simply propagate to infinity.

In the case of complete absorption, the deposited photon energy $\hbar \omega_\gamma$ is assumed large enough to heat the Bose system above $T_c$. True detection would then consist of a subsequent macroscopic flow or electrical resistance measurement that monitors whether or not absorption has taken place. Experimental realizations include superconducting nanowires \cite{NTH2012} and transition edge sensors \cite{IH2005}. Surrounding the microsample with appropriate filters also enables polarization or energy discrimination.

Implicit in the above is a classical outcome assumption: The photon is either absorbed or not, with probability given by the amplitude-magnitude-squared Born rule. To be consistent, it is argued that the equilibration dynamics subsequent to the initial absorption event (which may, e.g., kick a single boson into a higher energy state) is actually equivalent to such a measurement, collapsing the system wavefunction into one or the other macroscopic outcomes (superfluid plus escaping photon, or normal fluid plus EM vacuum), with classical probability derived from the Born rule. The collapse dynamics is controlled by the extremely rapid decoherence between the two boson final states \cite{foot:inel}.

The above conclusion is supported by a number of investigations \cite{Zurek1986,Zurek2019,W19} of quantum dissipation dynamics \cite{L1987,Mermin1991,SB2010}, generally based on the spin--boson or Caldeira--Leggett models of a localized spin or particle interacting with a harmonic oscillator bath. Here the incident photon may be thought of replacing the spin or particle, while the superfluid provides the bath. The collapse is argued to occur so quickly that decoherence of absorbed and escaping photon states is avoided, otherwise violating speed of light limitations. We emphasize that decoherence dynamics remains unitary, with the apparent random final state choice actually determined by details of the superfluid initial thermal state. For the absorption case, the information in the original pulse (detailed waveform and polarization), though unrecoverable, is fully encoded in the equilibrating superfluid state.

Given the above model of single photon detection, we next consider detection of multiple entangled photons. The aim is to recover the experimentally well-verified EPR predictions \cite{BellTest2001}. For space-like separation of (sufficiently rapid) detection events, individual detections are incapable of communicating, and it is argued that the result must be a ``cat state,'' i.e., a macroscopic superposition of permitted outcomes. However, the only way to detect such a state is to fully probe it, i.e., to enable sub-luminal communication between the measurement components. It is argued that such communication (as simple as illumination by the EM background) nucleates rapid decoherence, producing one or the other of the classical outcomes. One may think of separated light cones as defining the boundaries of two closed systems. Light cone intersection effectively converts the system from closed to open, enabling decoherence \cite{foot:Higgscollapse}.

In summary, the necessarily delayed (hence time-like) comparison of measurement results leads directly to wavefunction collapse, hence recovering EPR test results without appealing to superluminal communication. Other forms of single photon detection (e.g., intrinsically nonequilibrium systems such as avalanche diodes) require a different type of underlying physical model, but the key conclusions should still follow.

\paragraph{Model formulation:} The superfluid--EM system is described by the Hamiltonian $H = H_{BA} + H_\gamma$ where
\begin{eqnarray}
&&H_{BA} = \frac{1}{2m} \int_{V_B} d{\bf r}
\hat \psi^\dagger({\bf r}) (-i\nabla - q {\bf A})^2 \hat \psi({\bf r})
\nonumber \\
&&\ \ \
+\ \frac{1}{2} \int_{V_B} d{\bf r} \int_{V_B} d{\bf r}'
\hat \psi^\dagger({\bf r}) \hat \psi^\dagger({\bf r}')
V({\bf r}-{\bf r}') \hat \psi^\dagger({\bf r}') \hat \psi^\dagger({\bf r})
\nonumber \\
&&H_\gamma = \sum_{{\bf k},\alpha} \omega_\alpha({\bf k})
\hat a_{{\bf k}\alpha}^\dagger \hat a_{{\bf k}\alpha}
\label{1}
\end{eqnarray}
(adopting units where $\hbar = c = 1$) which couples the charge-$q$ boson (or Cooper pair) field operator $\hat \psi({\bf r})$, with standard commutation relation $[\hat \psi({\bf r}), \hat \psi^\dagger({\bf r}')] = \delta({\bf r}-{\bf r}')$, to the EM vector potential operator \cite{Baym}
\begin{equation}
{\bf A}({\bf r}) = \frac{1}{\sqrt{V}}
\sum_{\bf k} \sqrt{\frac{2\pi}{\omega_\alpha({\bf k})}}
\left(\hat a_{k\alpha} \hat {\bf e}_{{\bf k}\alpha} e^{i{\bf k} \cdot {\bf r}}
+ \hat a_{k\alpha}^\dagger \hat {\bf e}_{{\bf k}\alpha}^*
e^{-i{\bf k} \cdot {\bf r}} \right)
\label{2}
\end{equation}
with vacuum dispersion relation $\omega_\alpha({\bf k}) = c|{\bf k}|$, and polarization unit vectors $\hat {\bf e}_{{\bf k}\alpha}$, $\alpha = 1,2$ orthogonal to each other as well as to the wavevector ${\bf k}$. The photon creation and annihilation operators obey $[\hat a_{{\bf k}\alpha}, \hat a_{{\bf k}'\alpha'}^\dagger] = \delta_{{\bf k} {\bf k}'} \delta_{\alpha \alpha'}$.

The superfluid is confined to a microvolume $V_B$. Details of its state are not important for the present general treatment, but for tractable quantitative modeling the interaction potential $V({\bf r})$ might be viewed as weak, and the system treated as a near-ideal Bose gas. For a dilute Bose gas, one may use the effective potential $V({\bf r}) = v_0 \delta({\bf r})$ where $v_0 = 4\pi a_s$ with $a_s$ the $s$-wave scattering length, and the weakly interacting limit corresponds to $\rho_B a_s^3 \ll 1$, where $\rho_B = N_B/V_B$ \cite{FW1971,foot:wibgstate}. In addition, one is also free to work formally in the small $q$ limit so that boson--photon scattering is weak and can be treated within perturbation theory.

The EM model, on the other hand, exists in effectively infinite volume $V$, comprising a macroscopic laboratory setup, and enabling the plane wave decomposition (\ref{2}). For the entangled case we will consider multiple Bose systems $H^{(m)}_{BA}$, in separate volumes $V_B^{(m)}$, interacting with pulses that are physically separated but still governed by the same Hamiltonian $H_A$ and vector potential ${\bf A}$.

\paragraph{Single photon detection:} We treat first single photon detection. The initial state is of the form
\begin{equation}
|\Psi_0 \rangle = |\psi_{\alpha_0}({\bf k}_0) \rangle
\otimes |\psi_B^T \rangle
\label{3}
\end{equation}
in which $|\psi_B^T \rangle$ is a many body superfluid state, characterised by equilibrium temperature $T < T_c$, and the single photon state is $|\psi_{\alpha_0}({\bf k}_0) \rangle = \hat A_{{\bf k}_0 \alpha_0}^\dagger |\Omega_\gamma \rangle$ where $|\Omega_\gamma \rangle$ is the vacuum state and the operator
\begin{equation}
\hat A^{(C) \dagger}_{{\bf k}_0 \alpha_0} = \sum_{{\bf k},\alpha}
C_{{\bf k}\alpha}({\bf k}_0,\alpha_0) \hat n_{{\bf k}\alpha}^{-1/2}
\hat a_{{\bf k}\alpha}^\dagger
\label{4}
\end{equation}
adds a single photon with polarization $\alpha_0$ and a center wavevector ${\bf k}_0$ pointing towards the detector. The factor $\hat n_{{\bf k}\alpha}^{-1/2}$, with $\hat n_{{\bf k}\alpha} = \hat a_{{\bf k}\alpha}^\dagger \hat a_{{\bf k}\alpha}$, is unity for $|\Omega_\gamma \rangle$, but must be included for more general (e.g., thermal) populated background states. An example is the Gaussian
\begin{equation}
C_{{\bf k}\alpha}({\bf k}_0,\alpha_0) = {\cal N}({\bf k}_0,\Delta_0)^{-1/2}
e^{-|{\bf k} - {\bf k}_0|^2/4 \Delta_0^2} \delta_{\alpha,\alpha_0}
\label{5}
\end{equation}
with normalization ${\cal N}({\bf k}_0,\Delta_0) = \sum_{\bf k} e^{-|{\bf k} - {\bf k}_0|^2/2 \Delta_0^2} = \Delta_0^3 V/(2\pi)^{3/2}$, where we use $\sum_{\bf k} \to V \int d{\bf k}/(2\pi)^3$. The photon state is not in general an energy eigenstate, but has mean energy $E_{\alpha_0}({\bf k}_0) = \sum_{{\bf k},\alpha} \omega_\alpha({\bf k}) |C_{{\bf k}\alpha}({\bf k}_0,\alpha_0)|^2$.

The freely propagating photon state $e^{-it H_A} |\psi_{\alpha_0}({\bf k}_0) \rangle$
\begin{eqnarray}
|\psi_{\alpha_0}({\bf k}_0,t) \rangle
= \sum_{{\bf k} \alpha} C_{{\bf k}\alpha}({\bf k}_0,\alpha_0)
e^{i\omega_\alpha({\bf k}) t} |{\bf k},\alpha \rangle
\label{6}
\end{eqnarray}
corresponds to a propagating pulse: Using (\ref{5}), assuming $\Delta_0 \ll |{\bf k}_0|$, one obtains $\langle \Omega_\gamma| {\bf A}({\bf r}) |\psi_{\alpha_0}({\bf k}_0,t) \rangle \propto e^{-\Delta_0^2 |{\bf r} - {\bf v}^g_{\alpha_0}({\bf k}_0) t|^2}$ with group velocity ${\bf v}^g_\alpha({\bf k}) = \nabla_{\bf k} \omega_\alpha({\bf k}) = c \hat {\bf k}$. One may expect significant scattering or absorption only when the pulse center ${\bf r}_0(t) = {\bf v}^g_{\alpha_0}({\bf k_0}) t$ intersects $V_B$.

\paragraph{Post-interaction state:} The full evolution of the initial state is governed by the full Hamiltonian, $|\Psi(t) \rangle = e^{-it H} |\Psi_0 \rangle$. Based on conventional scattering physics, beyond some characteristic interaction time one might expect (mistakenly, as will be argued below) the form
\begin{eqnarray}
|\Psi(t) \rangle &=& C_0 |\psi_{\alpha_0}({\bf k}_0,t) \rangle
\otimes |\psi_B^T \rangle
+ C_T |\psi_\gamma^\mathrm{gnd} \rangle \otimes |\psi_B^{T'}(t) \rangle
\nonumber \\
&&+\ C_\mathrm{sc} |\Psi^\mathrm{sc}(t) \rangle.
\label{7}
\end{eqnarray}
The photon state in the first term corresponds to free propagation and leaves the superfluid state unchanged. The second term corresponds to full photon absorption, with $|\psi_B^T(t) \rangle$ an evolving excited Boson state, ultimately equilibrating to a thermally fluctuating state at a normal state temperature $T' > T_c > T$ \cite{foot:boseeq}. For a weakly focused photon, $\Delta_0^3 V_B \ll 1$, one expects $|C_T| \ll 1$. Unitary evolution ensures that the microstructure of $|\psi_B^{T'}(t) \rangle$ continues to encode for all time the full details of the original photon state, although this information will be irrecoverably lost among the boson modes. The final term $|\Psi^\mathrm{sc}(t) \rangle$ contains all remaining elastic and inelastic scattering contributions, in which the outgoing photon may have reduced energy and the superfluid volume is correspondingly partially excited. We will neglect this term in much of what follows, focusing on full absorption experimental outcomes. Note that both states $|\psi_B^{T,T'}(t) \rangle$ must also radiate, eventually equilibrating with the zero temperature EM vacuum. We will assume that this is a slow process and ignore it as well.

The problem with the state (\ref{7}) is that it is inconsistent with conventional measurement outcome predictions, which must produce a definite bulk superfluid or normal fluid response. Specifically, the associated density matrix $\hat \rho(t) = |\Psi(t) \rangle \langle \Psi(t)|$ is a pure state, with both diagonal and off-diagonal terms. However, the measurement result should be governed by the macroscopic mixed state density matrix
\begin{eqnarray}
\hat \rho_\mathrm{meas}(t) &=& p_0
(|\psi_{\alpha_0}({\bf k}_0,t) \rangle
\langle \psi_{\alpha_0}({\bf k}_0,t)|)
\otimes (|\psi_B^T \rangle \langle \psi_B^T|)
\nonumber \\
&&+\ p_T (|\Omega_\gamma \rangle
\langle \Omega_\gamma|)
\otimes (|\psi_B^{T'}(t) \rangle \langle \psi_B^{T'}(t)|)
\nonumber \\
&&+\ p_\mathrm{sc} |\Psi^\mathrm{sc}(t) \rangle
\langle \Psi^\mathrm{sc}(t)|,
\label{8}
\end{eqnarray}
where $p_0 = |C_0|^2$, $p_T = |C_T|^2$, $p_\mathrm{sc} = |C_\mathrm{sc}|^2$ are now classical (Born rule) probabilities for each outcome. The last term in general will actually decompose into a further sum of distinct final scattering states. The problem is that it is not physically possible for a measurement producing the absorbed photon outcome to somehow reach out, perhaps significantly after the fact, to eliminate the escaping photon in the unperturbed superfluid outcome \cite{foot:spad}. Once the photon pulse has left the vicinity of the superfluid microsystem, the EM state is governed entirely by $H_A$ which cannot create or destroy photons.

The alternative, argued for here, is that the state $|\Psi(t) \rangle$ must itself produce the classical measurement outcome as part of the photon absorption process, with unitary evolution toward one of the states in (\ref{8}), and with effectively random choice (according to the above probabilities) governed by detailed equilibrium dynamics of the initial state $|\psi_B^T \rangle$. This outcome, enabled by the macroscopic nature of the boson state which effectively continuously measures itself, is supported by a long history of quantum decoherence investigations \cite{Zurek1986,Zurek2019,W19,L1987,Mermin1991}). In particular, macroscopic thermal decoherence times on the order of $10^{-15}$ s or shorter  \cite{Zurek1986,W19}, between the superfluid state and the developing normal fluid state during and immediately following photon interaction (as the number of superfluid particles impacted by the absorption grows to macroscopic size) is short compared to the ($10^{-12}$ s, say) photon travel time through $V_B$. The scattering state (\ref{7}) does not even have the opportunity to form, and one converges instead to one of the ``pointer state'' outcomes \cite{Zurek2019} entirely during the absorption process. It would be interesting to explore details of the latter within the weakly interacting model (\ref{1}). Note that the subsequent thermodynamic measurement is redundant here, playing no role in the formation of the final state. It serves only to display the result to the observer.

\paragraph{Multi-photon states:} Details of decoherence/wavefunction collapse for multiphoton detection are not critical here. We assume only that separated detection processes act independently according to the previous analysis. We consider only two photon states. Generalization to higher numbers is straightforward.

The simplest two-photon states are of the form
\begin{equation}
|\psi_{\alpha_1 \alpha_2}({\bf k}_1,{\bf k}_2) \rangle
= \hat A^{(C) \dagger}_{{\bf k}_1 \alpha_1}
\hat A^{(C) \dagger}_{{\bf k}_2 \alpha_2}
|\Omega_\gamma \rangle.
\label{9}
\end{equation}
We assume distinct pulses: the center wavevectors ${\bf k}_1,{\bf k}_2$ are sufficiently separated that their pulse supports do not significantly overlap, e.g., $|{\bf k}_1 - {\bf k}_2| \gg \Delta_0$ for Gaussian pulses (\ref{5}), and the directions $\hat {\bf k}_{1,2}$ point toward separate superfluid detectors. The combined initial state is
\begin{equation}
|\Psi_0^{(2)} \rangle = |\psi_{\alpha_1 \alpha_2}({\bf k}_1,{\bf k}_2) \rangle
\otimes |\psi_{B,1}^{T_1} \rangle \otimes |\psi_{B,2}^{T_2} \rangle
\label{10}
\end{equation}
with in general distinct temperatures $T_1,T_2 < T_c$.

The product states (\ref{9}) are the simplest entangled states. More interesting are superpositions of the form
\begin{equation}
|\psi_D^{(2)} \rangle
= \sum_{\alpha_1, \alpha_2} D_{\alpha_1 \alpha_2}
|\psi_{\alpha_1 \alpha_2}({\bf k}_{1,\alpha_1}, {\bf k}_{2,\alpha_2}) \rangle,
\label{11}
\end{equation}
superposing states with different polarizations and center wavevectors. For simplicity, we point the four wavevectors ${\bf k}_{l,\alpha}$, $l=1,2$, $\alpha=1,2$ towards four different detectors. The usual antisymmetric state corresponds to $D^A_{\alpha_1 \alpha_2} = \frac{1}{\sqrt{2}}
(-1)^{\alpha_2} \delta_{\alpha_1,\bar \alpha_2}$ in which $\bar \alpha = 3-\alpha$ is the orthogonal polarization. The combined initial state is
\begin{equation}
|\Psi_0^{(D)} \rangle  = |\psi_D^{(2)} \rangle
\otimes |\psi_{B,1}^{T_1} \rangle
\otimes |\psi_{B,2}^{T_2} \rangle
\otimes |\psi_{B,3}^{T_3} \rangle
\otimes |\psi_{B,4}^{T_4} \rangle.
\label{12}
\end{equation}

\paragraph{Two photon detection:} Considering first the product state (\ref{9}), assuming independent scattering and absorption events at each detector, the density matrix effectively becomes a direct product of outcomes in (\ref{8}) for each detector: the wavefunction collapse step trivially recovers the Born rule probabilities for each independent classical outcome. For example, the key two-photon detection term takes the form
\begin{eqnarray}
\delta \rho^{(2)}_\mathrm{meas}(t) &=& p^{(1)}_T p^{(2)}_T
(|\Omega_\gamma \rangle \langle \Omega_\gamma|)
\otimes\ (|\psi_{B,1}^{T_1'}(t) \rangle \langle \psi_{B,1}^{T_1'}(t)|)
\nonumber \\
&&\otimes\ (|\psi_{B,2}^{T_2'}(t) \rangle \langle \psi_{B,2}^{T_2'}(t)|)
\label{13}
\end{eqnarray}
determined by the product of absorption probabilities.

\paragraph{Entangled photon detection and cat states:} Proceeding to the entangled superposition state (\ref{12}), linearity allows us to treat each term independently, generating the superposition of scattering states:
\begin{equation}
|\Psi^{(D)}(t) \rangle = \sum_{\alpha_1, \alpha_2}
D_{\alpha_1 \alpha_2} |\Psi^{(2)}_{\alpha_1 \alpha_2}(t) \rangle,
\label{14}
\end{equation}
in which each $|\Psi^{(2)}_{\alpha_1 \alpha_2}(t) \rangle$ corresponds, \emph{initially,} to independent time evolution of the simple photon product states (\ref{10}), appropriate to the different pulse parameters. Thus, each state randomly produces one of the pointer state outcomes \cite{Zurek2019} encoded in (\ref{8}), and (\ref{14}) is the \emph{quantum superposition} of such macroscopic states. For example, focusing on the antisymmetric entangled case, and for the outcome in which exactly two photons are detected, the final state (which will occur with a suitable product of classical probabilities) is the superposition $\frac{1}{\sqrt{2}}(|\psi^{(D)}_1(t) \rangle - |\psi^{(D)}_2(t) \rangle)$ with components
\begin{eqnarray}
|\psi^{(D)}_1(t) \rangle &=& |\Omega_\gamma \rangle
\otimes |\psi_{B,1}^{T_1'}(t) \rangle \otimes |\psi_{B,2}^{T_2} \rangle
\nonumber \\
&&\otimes\ |\psi_{B,3}^{T_3'}(t) \rangle \otimes |\psi_{B,4}^{T_4} \rangle
\nonumber \\
|\psi^{(D)}_2(t) \rangle &=& |\Omega_\gamma \rangle
\otimes |\psi_{B,1}^{T_1} \rangle \otimes |\psi_{B,2}^{T_2'}(t) \rangle
\nonumber \\
&&\otimes\ |\psi_{B,3}^{T_3} \rangle \otimes |\psi_{B,4}^{T_4'}(t) \rangle.
\label{15}
\end{eqnarray}
The first corresponds to detection of photons ${\bf k}_{1,\alpha_1}$ and ${\bf k}_{2, \alpha_2}$ in detectors 1 and 3, respectively, while the second corresponds to detection of photons ${\bf k}_{1,\alpha_2}$ and ${\bf k}_{2,\alpha_1}$ in detectors 2 and 4, respectively.

\paragraph{Cat state collapse:} The independence assumption, hence survival of such a cat state, requires that the detectors do not interact. This certainly is the case for space-like absorption event separation. However, as time progresses (e.g., within $\sim 10$ ps for a typical lab setup), given that these states differ macroscopically (superfluid vs.\ normal for different subsets of detectors, or perhaps more subtly for inelastic detection outcomes), the conventional picture implies that (\ref{15}) should decohere to form the diagonal density matrix:
\begin{equation}
\hat \rho_{1,2}(t) = \frac{1}{2} |\psi^{(D)}_1(t) \rangle \langle \psi^{(D)}_1(t)|
+ \frac{1}{2} |\psi^{(D)}_2(t) \rangle \langle \psi^{(D)}_2(t)|,
\label{16}
\end{equation}
corresponding to definite classical outcomes with perfect EPR polarization anticorrelation: $\alpha_1$ at detector 1, $\bar \alpha_2$ at detector 3; $\alpha_2$ at detector 2, $\bar \alpha_1$ at detector 4.

Modeling this final collapse depends critically on the system details and its surroundings \cite{foot:widesep}. Here we summarize key features of such a model, and explore the resulting dynamics using a very simple effective model motivated by entangled qubit decoherence \cite{W19}. One may imagine an EM environment, whether passive (e.g., thermal background) or active (e.g., deliberate EM interrogation), illuminating the detectors, with weak photon reflections between them, generating (sub-luminal time scale) interactions between their states.

As the simplest possible description, we construct a high level superspin-$\frac{1}{2}$ model of the two possible combined detector states (\ref{15}), with spin states $|\psi^{(D)}_1 \rangle \to |+ \rangle$ and $|\psi^{(D)}_2 \rangle \to |- \rangle$, and associated spin operator $\hat {\bm \Sigma}$ with $\hat \Sigma_z |\pm \rangle = \pm |\pm \rangle$. The simplest environment interaction might be described by a linear (spin--boson) coupling, with Hamiltonian
\begin{equation}
H_{SB} =  \sum_l \omega_l \hat a_l^\dagger \hat a_l - {\bf h} \cdot \hat {\bm \Sigma}
- \hat \Sigma_z \sum_l \lambda_l (\hat a_l + \hat a_l^\dagger),
\label{17}
\end{equation}
in which the EM background is subsumed into a set of effective high level boson modes $\hat a_l$, with frequencies $\omega_l$ (reflecting, e.g., the longer wavelength part of the photon spectrum most sensitive to the bulk detector state) and coupling constants $\lambda_l$ (quantifying interaction of these modes with the detectors). In addition to an average magnetic field ${\bf h}$, the mode displacement $\hat q_l = \frac{1}{\sqrt{2}}(\hat a_l + \hat a_l^\dagger)$ acts as a fluctuating magnetic field coupled to the spin states $\hat \Sigma_z = \pm 1$, with all internal structure of these states now ignored. The field $h_z$ encodes any bias between the states $|\pm \rangle$, reflecting perhaps any physical asymmetry within the detector setup. Here we will take $h_z = 0$, required to obtain the equal probabilities in (\ref{16}). The $x,y$ components generate tunneling between the states, e.g., $\hat \Sigma_x |\pm \rangle = |\mp \rangle$, and bias the system towards the eigenstates of ${\bf h} \cdot \hat {\bm \Sigma}$ (cat state superpositions of $|\pm \rangle$). Choosing spin-coordinates to set $h_y = 0$, one may then estimate $h_x \approx \langle \psi^{(D)}_2| H |\psi^{(D)}_1 \rangle$ from the tunneling matrix element between the two states generated by the full microscopic Hamiltonian. This matrix element physically reflects the dynamics generated by $H$ leading to EM interaction between the two states, and is generally nonzero (though, in some cases it may require a second or higher order calculation). Although the microscopic coupling may be very weak, the macroscopic difference between the states can still lead to the high tunneling rate (e.g., sub-ps period \cite{W19}) required for rapid decoherence.

Properties of the model (\ref{17}) are well known \cite{L1987,Mermin1991,SB2010,W19}, and we summarize here only the key consequences. The spin--boson coupling is quantified by the spectral function $J(\omega) = \pi \sum_l \lambda_l^2 \delta(\omega-\omega_l)$. The model form $J(\omega) = 2\pi \alpha \omega^s e^{-\omega/\omega_c}$, with high frequency cutoff $\omega_c$ and low frequency power law exponent $s$ (depending, e.g., on boson dispersion relation and scattering amplitude scaling), characterizes a number of physical situations. For $s \leq 1$ the interaction induces a quantum phase transition to a ferromagnetic state with nonzero $M_z = \langle \hat \Sigma_z \rangle \sim (\alpha - \alpha_c)^{\beta(s)}$ with increasing $\alpha > \alpha_c(h_x) \sim h_x^2/\omega_c^{s+1}$ and order parameter exponent $\beta(s)$ \cite{foot:qpt}. So long as the tunneling bias $h_x$ is small, i.e., $\alpha/\alpha_c(h_x) \gg 1$, which one generally expects on physical grounds, one will also be in the regime $|M_z| \simeq 1$. The sign of $M_z$ is again selected randomly, through details of the initial photon background state, and corresponds precisely to the choice of collapse to one or the other detection state $|\psi^{(D)}_{1,2} \rangle$, consistent with experimental observation \cite{BellTest2001}.

\paragraph{Conclusions:} The key consequence of the model (\ref{17}), despite its simplicity, is the confirmation that fluctuating environmental ``observations'' of detector cat states (which, based only on properties of the many body Schr{\"o}dinger equation, we have argued must accompany space-like entangled photon detections) are sufficient to rapidly drive the preference for one or the other component ``classical'' state as they return to causal contact. The work here also clarifies the key role of causal separation in permitting temporary survival of such states: rapid collapse will occur absent extreme isolation, e.g., that underlying error-free quantum computation.

Interesting future work would include quantitative estimation of the parameters in (\ref{17}) (or that of a more general model, if needed) based on the microscopic model (\ref{1}) for various measurement scenarios, including speed of light delay effects entering the $\lambda_l$. More detailed consequences, especially estimates of wavefunction collapse times, should be explored as well. Taking a step back, the density matrix (\ref{8}) for the single photon detection scenario underlies all other scenarios. Its form is completely conventional, underlying numerous experimental results. Its detailed derivation from (\ref{1}), within the various weak interaction assumptions required to obtain a tractable problem, would be extremely interesting.


\begin{thebibliography}{}

\bibitem{NTH2012} C. M. Natarajan, M. G. Tanner, and R. H. Hadfield,
\href{https://doi.org/10.1088/0953-2048/25/6/063001}{Superconductor Sci.\ Tech.\ \textbf{25}, 063001 (2012)}. \href{https://arxiv.org/abs/1204.5560}{arXiv:1204.5560}.

\bibitem{IH2005} K. D. Irwin and G. C. Hilton,
\emph{Cryogenic Particle Detection,} pp.\ 81--97 (Springer, 2005).

\bibitem{foot:inel} Classical outcomes occur also through decoherence of the macroscopic states associated with the various inelastic processes. These are neglected here to simplify the discussion.

\bibitem{Zurek1986} W. H. Zurek,
NATO ASI \textbf{B135}, 145--149, G. T. Moore and M. O. Scully, eds. (Plenum, 1986). \href{https://arxiv.org/abs/quant-ph/0302044}{arXiv:quant-ph/0302044 [quant-ph]}.

\bibitem{Zurek2019} W. H. Zurek,
\href{https://doi.org/10.1098/rsta.2018.0107}{Phil.\ Trans.\ R.\ Soc.\ A \textbf{376}: 20180107 (2018)}. \href{https://arxiv.org/abs/1807.02092}{arXiv:1807.02092 [quant-ph]}.

\bibitem{W19} P. B. Weichman, ``A quantum phase transition implementation of quantum measurement,'' (arXiv link, 2019).

\bibitem{L1987} A. J. Leggett, S. Chakravarty, A. T. Dorsey, M. P. A. Fisher, A. Garg, and W. Zwerger,
\href{https://doi.org/10.1103/RevModPhys.59.1}{Rev.\ Mod.\ Phys.\ \textbf{59}, 1--85, (1987)}.

\bibitem{Mermin1991} N. D. Mermin,
\href{https://doi.org/10.1016/0378-4371(91)90201-M}{Physica A \textbf{177}, 561--566 (1991)}.

\bibitem{SB2010} For a review, see, e.g., K. Le Hur,
Chapter 8 in \href{https://www.crcpress.com/Understanding-Quantum-Phase-Transitions/Carr/p/book/9781439802519} {\emph{Understanding Quantum Phase Transitions}}, edited by L. D. Carr (CRC Press, Taylor \& Francis, 2010).

\bibitem{BellTest2001} M. A. Rowe, D. Kielpinski, V. Meyer, C. A. Sackett, W. M. Itano, C. Monroe, and D. J. Wineland,
\href{https://doi.org/10.1038/35057215}{Nature \textbf{409}, 791--794 (2001)}.

\bibitem{foot:Higgscollapse} An extreme version of this is the expanding universal destruction of a metastable Higgs boson vacuum: See, e.g., A. Kusenko,  \href{https://physics.aps.org/articles/v8/108}{Viewpoint: Are We on the Brink of the Higgs Abyss?} Separated measurement decoherence appears to be tame in comparison.

\bibitem{Baym} G. Baym, \emph{Lectures on Quantum Mechanics} (Benjamin/Cummings, 1969).

\bibitem{FW1971} A. L. Fetter and J. D. Walecka, \emph{Quantum Theory of Many Particle Systems} (McGraw-Hill, 1971), especially Chap.\ 14.

\bibitem{foot:wibgstate} For example, in the weakly interacting equilibrium limit one obtains superfluid state $|\psi_B^T \rangle_\mathrm{eq} = |N_0 \rangle_0 \otimes_{{\bf p} \neq 0} |n^T_{\bf p} \rangle_{\bf p}$ with macroscopic condensate number $N_0 = n_0 V_B = N_B - \sum_{{\bf p} \neq 0} n^T_{\bf p}$ . The momentum state occupation numbers $n^T_{\bf p}$ are Poisson distributed with mean $n^T_B(E_{\bf p})$ defined by the Bose occupation factor $n^T_B(x) = (e^{x/T} - 1)^{-1}$ and Bogoliubov energy $E_{\bf p} = \sqrt{\varepsilon_{\bf p}(\varepsilon_{\bf p} + 2 n_0 v_0)}$ with $\varepsilon_{\bf p} = {\bf p}^2/2m$ \cite{FW1971}. In the normal state $n_0 = 0$ and the mean occupancies become $n^{T'}_B(\varepsilon_{\bf p} - \mu)$ with chemical potential $\mu$ enforcing  $\sum_{\bf p} n^{T'}_B(\varepsilon_{\bf p} - \mu) = N_B$.

\bibitem{foot:boseeq} For weak scattering, the $-i (q/m) \psi^\dagger {\bf A} \cdot \nabla \psi$ interaction has the leading effect of promoting a single Boson from energy $\varepsilon_{\bf p} \to \varepsilon_{\bf p} + \hbar \omega$. Following this, neglecting further EM interactions, the system equilibrates to the state $|\psi_B^{T'} \rangle_\mathrm{eq}$ through boson--boson scattering.

\bibitem{foot:spad} An effect often described as ``spooky action at a distance.''

\bibitem{foot:widesep} This is especially the case for widely separated measurements where other macroscopic events, such as observer reaction, could occur prior to light cone intersection, and would need to be included in the wavefunction collapse.

\bibitem{foot:qpt} Strictly speaking, the ordered phase exists only at $T=0$, but for $T > 0$ not too large the tunneling between the two macroscopic phases that destroys the order will be unobservable on experimental time scales.

\end{thebibliography}
\end{document}